\documentclass{nature}

\usepackage{graphicx}
\usepackage{threeparttable}
\usepackage{hyperref}
\usepackage{color}
\usepackage{subfig}
\usepackage{caption}

\usepackage{amssymb}
\usepackage{graphicx}
\usepackage{color}
\usepackage[normalem]{ulem}
\usepackage{soul}

\newcommand{\xmm}{{XMM-Newton} }

\newcommand{\swift}{{Swift} }
 
\newcommand{\arcsec}{\mbox{$^{\prime\prime}$}}

\newcommand{\nh}{cm$^{-2}$}

\newcommand{\apj}{{\it Astrophys. J.}}

\newcommand{\mnras}{{\it Mon. Not. R. Astron. Soc.}}
\newcommand{\apjl}{{\it Astrophys. J. Let.}}
\newcommand{\aap}{{\it Astron. Astrophys.}}

\newcommand{\nat}{{\it Nature}}
\newcommand{\na}{New Astron. Rev.}
\newcommand{\sci}{{\it Science}}

\newcommand{\ergs}{${\rm erg \ cm^{-2} \ s^{-1}}$ }
\newcommand{\erg}{${\rm erg \ s^{-1}}$ }
\newcommand {\um}{$\mu$m~}

\def\ts     {\thinspace}

\def\kms    {\ifmmode{{\rm \ts km\ts s}^{-1}}\else{\ts km\ts s$^{-1}$}\fi}
\def\msol   {\ifmmode{{\rm M}_{\odot}}\else{M$_{\odot}$}\fi}
\def\lsol   {\ifmmode{{\rm L}_{\odot}}\else{L$_{\odot}$}\fi}
\def\zsol   {\ifmmode{{\rm Z}_{\odot}}\else{Z$_{\odot}$}\fi}

\def\ltsima{$\; \buildrel < \over \sim \;$}
\def\simlt{\lower.5ex\hbox{\ltsima}}
\def\gtsima{$\; \buildrel > \over \sim \;$}
\def\simgt{\lower.5ex\hbox{\gtsima}}

\newcommand{\msun}{{\rm\,M$_\odot$}}

\newcommand{\src}{{\rm\,OGLE16aaa~}}
\newcommand{\srcs}{{\rm\,OGLE16aaa}}

\usepackage{floatrow}
\DeclareFloatFont{tiny}{\scriptsize}
\title
{X-ray flares from the stellar tidal disruption by a candidate supermassive black hole binary}

\newcounter{firstbib}

\author{Xinwen~Shu$^{1*}$, Wenjie~Zhang$^{1}$, Shuo~Li$^{2}$, Ning~Jiang$^{3}$, Liming~Dou$^{4}$, Zhen~Yan$^{5}$, Fu-Guo~Xie$^{5}$, Rongfeng Shen$^{6}$, Luming~Sun$^{1}$, Fukun~Liu$^{7, 8}$, Tinggui~Wang$^{3}$  } 
\begin{document}

\maketitle

\let\thefootnote\relax\footnote{
\begin{affiliations}
\item{Department of Physics, Anhui Normal University, Wuhu, Anhui 241002, China. 
	{\bf $^{*}$email: xwshu@ahnu.edu.cn}}
\item{National Astronomical Observatories, Chinese Academy of Sciences, Beijing 100101, China} 
\item{CAS Key Laboratory for Researches in Galaxies and Cosmology, Department of Astronomy, 
      University of Science and Technology of China, Hefei, Anhui 230026, China}
\item{Center for Astrophysics, Guangzhou University, Guangzhou 510006, China}
\item{Shanghai Astronomical Observatory, Chinese Academy of Sciences, Shanghai 200030, China}
\item{School of Physics and Astronomy, Sun Yat-Sen University, Zhuhai 519082, China}
\item{Department of Astronomy, Peking University, Beijing 100871, China}
\item{Kavli Institute for Astronomy and Astrophysics, Peking University, Beijing 100871, China}
\end{affiliations}
}




\hfill

\begin{abstract}
Optical transient surveys have led to the discovery of dozens of stellar tidal disruption events (TDEs) by massive black hole in the centers of galaxies. 
Despite extensive searches, X-ray follow-up observations have produced no or only weak X-ray detections 
in most of them. 
Here we report the discovery of delayed X-ray brightening around 140 days after the optical outburst in the 
TDE OGLE16aaa, followed by several flux dips during the decay phase. These properties are unusual for standard TDEs and could be explained by the presence of supermassive black hole binary or patchy obscuration. In either scenario, the X-rays can be produced promptly after the disruption but are blocked in the early phase, possibly by a radiation-dominated ejecta which leads to the bulk of optical and ultraviolet emission. Our findings imply that the reprocessing is important in the TDE early evolution, and X-ray observations are promising in revealing supermassive black hole binaries. 
\end{abstract}


\section{Introduction}

Almost all massive galaxies appear to contain a central supermassive black hole (SMBH) with 
a mass of $\simgt$$10^6$\msun~(where \msun~refers to the solar mass)\cite{Graham2016}, yet most of them 
remain unobservable due to the lack of enough radiative output 
through accretion process. 
Stars that pass sufficiently close to a SMBH can be torn apart when the
tidal force of the SMBH exceeds its self-gravity. 
While roughly half of the stellar material will be ejected, the other half will 
remain bound and eventually be accreted, producing 
a luminous flare of electromagnetic radiation\cite{Rees1988}. 
These tidal disruption events (TDE) 
not only provide novel means of probing SMBH in otherwise 
quiescent galaxies\cite{Komossa2015}, but also serve as a unique laboratory 
for studying the formation and evolution of accretion disk (e.g., \cite{Piran2015, Gezari2017, Wevers2019}), 
the launch of relativistic jets\cite{Burrows2011, Mattila2018}, 
as well as gravitational wave (GW) emission through coalescence of SMBH binaries\cite{Liu2014, Hayasaki2016}. 

Early theoretical works predict that TDEs should produce a bright
thermal emission {peaking mainly in soft X-ray bands}, 
which originates from a newly formed accretion disk\cite{Rees1988, Ulmer1999}. 
{The effective temperature of thermal radiation produced by disk accretion\cite{Miller2015} 
is $T_{\rm eff} \approx 4\times10^5$ $M_{\rm 6}^{-1/4}$K, where $M_{\rm 6}=M_{\rm BH}/(10^6)$\msun.}
The TDEs discovered {in the ultraviolet (UV) and optical bands}, however, are found to have surprisingly  
low blackbody temperature (of $\sim1-4\times10^4$ K) and correspondingly large black 
body radii ($\sim10^{14-15}$ cm)\cite{Hung2017}
, which are difficult to reconcile with the predicted radiation from a compact accretion disk\cite{Ulmer1999, Komossa2015}. 
The {UV}/optical emission can be instead powered by shocks from stream self-collisions during the 
formation of disk\cite{Piran2015, Bonnerot2017}, or conversely thermal reprocessing of accretion power 
by a layer of gas at large radii\cite{Strubbe2009,Metzger2016}.  
X-ray observations within {the first few months} of discovery are critical to disentangle 
the dominant emission mechanisms of optical light, yet only a handful of optical TDEs 
have been successfully detected with X-ray emission\cite{van2016, Gezari2017, Wevers2019, van2020}. 
The origin of dominant {UV}/optical emission and its association with {the X-ray one} in TDEs still remains unclear. 



{Here we show 
delayed X-ray brightening by a factor of $>$60, approximately 140 days after the optical flare, in the TDE \srcs, 
followed by several dips of X-ray emission during the afterwards decay phase.    
These properties are unusual among standard TDEs and are instead consistent with either the tidal disruption 
by SMBH binary or changes in absorption along the line of sight. In this context, the X-ray non-detections at the very 
early times could be due to obscuration, in which the reprocessed accretion radiation may power the bulk of observed UV and optical emission. 
}

\section{Results}

     
 {\bf \noindent Optical and UV light curve analysis.} 
 The optical transient, \src ($\rm RA_{J2000}=$ 01h07m20.88s, $\rm DEC_{J2000}=$ -64deg16'20.7''), 
 was discovered by the Optical Gravitational Lensing Experiment (OGLE-IV, \cite{Udalski2015}), 
 {and its Transient Detection System\cite{Wyrzykowski2014}} 
 on 2 Jan 2016 at a redshift of $z=0.1655$, coincident with the nucleus of its 
 host galaxy\cite{Wyrzykowski2017}.  
 The I-band light curve shows a rise over $\sim$30 rest-frame days of about 3 mag 
 above the quiescent brightness of the host galaxy, {reaching} a peak I-band magnitude 
 of 18.98 mag on 20 Jan 2016, and {subsequently declining by $\sim0.6$ mag (to I = 19.57 mag) over two weeks.  
 Then its flux starts to increase again at $t\approx20$ days with respect to the initial peak, 
 {reaching} a secondary maximum of I = 19.33 mag within one week, with evidence for another decline until 
 $t=50$ days (Figure 1a). After a time interruption in the optical observations, 
 the light curve appears to show a plateau since $t=146$ days, with 
 a flux comparable to that of pre-flare phase, suggestive of the dominance of the host emission. 
 Following the initial optical peak,} \src was monitored at three UV bands 
 (UVW2, $\lambda_{\rm eff}=1928$\AA; UVM2, $\lambda_{\rm eff}=2246$\AA; 
 UVW1, $\lambda_{\rm eff}=2600$\AA)
 {by the Ultraviolet and Optical Telescope (UVOT)\cite{Roming2005}, as well as the 
 {X-ray band} by the X-ray Telescope (XRT)\cite{Burrows2005} on board the \swift observatory\cite{Gehrels2004}} (Supplementary Table 1). 
 We have analysed all publicly available \swift observations 
 {(see Methods: X-ray and UV data reduction).} 
 By including more data after March 2016, results of our reanalysis 
 are generally consistent with those reported in the literature\cite{Wyrzykowski2017}, 
 and confirm that the UV emission decayed as expected from a TDE. 
 {While the luminosity evolution can be described by the canonical $t^{-5/3}$ power model, 
 the exponential decline model ($L=L_0e^{-(t-t_0)/\tau}$) is able to fit the data equally well (Supplementary Figure 2). 
 Since the UV emission appears to decay to a plateau level, we can not distinguish between the 
 two models with current observations. 
 Note that there is a tentative evidence for the UVW2 and UVM2 bands displaying a similar 
 re-brightening as the optical. However, it is not statistically significant (Supplementary Note 2). 
 }
 In comparison with {the {GALEX}\cite{Martin2005} }data taken on 2003, 
 the UV emission at $\sim2300$\AA~(NUV)~increased by a factor of 7.7, reaching a 
 recorded peak luminosity as high as $10^{44}$\erg.  


 {\bf \noindent X-ray flux evolution.} 
 Despite frequently monitored by the XRT, no X-ray emission was detected in either individual 
 or stacking observations in early times from 19 Jan to 8 June 2016 {(Figure 1b)}, with an 0.3-2 keV luminosity 
 $L_{\rm X}<8.9\times10^{41}$\erg ($3\sigma$ upper limit, assuming a blackbody spectrum of 
 temperature of $kT_{\rm BB}=60$ eV, see the spectral analysis below). Hereafter, we defined the non-detection 
 as less than one net count, or the probability of having source counts from the background 
 greater than 0.05 in Poisson statistics, {and accordingly reported three net counts as the $3\sigma$ upper limit 
 on the flux of interest}. The source was first detected by {deeper {XMM-Newton}\cite{Jansen2001}} observation 
 on 9 June 2016 {(XMM1)}, $\sim$141 days later since the optical peak, with $L_{\rm X}=2.9\times10^{42}$\erg. 
 About a week later, \swift caught the source at an even higher luminosity, {reaching} an 
 X-ray peak at $L_{\rm X}\sim7\times10^{43}$\erg between 17 June and 21 June 2016. 
 The luminosity then decreased by more than an order of magnitude in the following \xmm and 
 \swift observations. 
 {Details of X-ray observations are shown in Table 1.}
 
 {
 The overall long-term evolution of the X-ray luminosity is unique among known X-ray detected 
 TDEs, in particular the fast-rise to the peak within only two weeks. 
   Another striking feature in the X-ray lightcurve is that the source became completely 
   invisible to \swift on 2016 Nov 17 after the peak, with a $3\sigma$ upper limit on the flux of 
   $9.4\times10^{-14}$\ergs in the 0.3--2 keV (or equally $L_{\rm X}<7.2\times10^{42}$\erg). Then it recurred to a much higher 
   flux level in the second XMM observation on 2016 Nov 30 (XMM2), followed by a new flux dip. 
   The time scale of the flux increase (about two weeks) is somewhat consistent with that 
   observed in the earlier brightening epoch at $\sim$140 days.  
   We argue that the \swift non-detection before XMM2 is significant. 
   Assuming mild source variability in standard TDE evolution and so the same flux as that obtained with the XMM2 observation, 
   we would expect around $10$ counts in the 0.3--2 keV to be detected by \swift for an exposure of 1.7 ks. 
   For a Poisson distribution, the chance of detecting less than one photon in the \swift observation is 
   $5\times10^{-5}$. Conversely, we detected $\sim$5-9 net counts in two subsequent \swift observations. 
   The probability of having these counts from the background 
   is $<$0.15\% in Poisson statistics (or at a significance level of 99.85\%). 
   Note that previous study of this event\cite{Wyrzykowski2017, Auchettl2017} either 
   failed to detect the X-ray emission at early times (around optical flare), or used only the 
   partial observations from the \xmm and \swift datasets to describe the X-ray evolution up to 
   17 Nov 2016 ($\sim$300 days since optical flare). 
   {Our results are broadly consistent with the recent study\cite{Kajava2020}, confirming 
   the rapid X-ray brightening in \srcs. However, \cite{Kajava2020} does not identify }
   further flux flares and dips in the decay phase,     
   which are crucial to lead to the advanced interpretation of overall 
   observing properties of the source that will be presented in the next Section. 
   
   }

 {\bf \noindent X-ray spectral analysis.} 
 Another notable property for the source is the quasi-soft X-ray spectra 
 that {lack} emission above $\sim$2 keV, typical for TDEs detected in the {X-ray bands}. 
 This is most evident in the two \xmm observations that have the best spectral 
 quality. {The spectrum obtained from the first \xmm observation can be well fitted with a 
 single blackbody component ({zbbody} in XSPEC) with a temperature of $kT_{\rm BB}=58\pm5$ eV, modified by 
 the Galactic absorption ($N_{\rm H}=2.7\times10^{20}$ cm$^{-2}$), as shown in {Figure 2a}.  
 No additional absorption intrinsic to the source was required. 
 We used the same single blackbody model to describe the X-ray spectrum in the second \xmm observation, 
 and found a similar temperature for the blackbody emission (Figure 2c). 
 The data to model ratios, however, show a clear excess of emission at energies above 
 $\sim$0.7 keV {(Figure 2b)}, suggesting that the single blackbody model is not enough 
 to describe the X-ray spectrum of XMM2. 
 The spectral fitting result is improved significantly by the addition of a powerlaw component 
 to the above model {(Figure 2f)}. 
 The overall $\chi^2$ decreased by 46 for two extra parameters, with a $F$-test probability of 
 $3.6\times10^{-9}$. 
 In this case, we obtain a best-fitted blackbody temperature of 
 $kT_{\rm BB} = 60^{+9}_{-11}$ eV, which is still comparable to that obtained from XMM1. } 
 Albeit with large uncertainties, the additional power-law
 component is steep with a photon index $\Gamma=6.4\pm0.5$, that 
 is rarely seen in TDEs\cite{Auchettl2017}. 
 Alternatively, the excess emission can be described by another  
 blackbody component with a higher temperature of $kT_{\rm BB} = 90^{+20}_{-13}$ eV 
 {{(Figure 2d and 2e)}}. In this case, the temperature 
 for the primary blackbody component decreases slightly to $kT_{\rm BB} = 51^{+5}_{-8}$ eV. 
 {Results of X-ray spectral analysis are presented in Table 2.}
 Similar spectral analysis was also performed for the \swift data. 
 Unfortunately the spectral signal-to-noise (S/N) ratios are not sufficient for a 
 precise determination of the temperature even after stacking the data from individual observations. 
 In the Supplementary {Figure 3}, we present the combined X-ray spectrum obtained by 
 \swift at the peak. 
 The majority of the counts for the combined spectrum fall into the low-energy range of 0.3-0.7 keV, 
 indicating the spectrum has remained soft. 
 The spectrum can be described by a blackbody model with $kT_{\rm BB} = 73^{+22}_{-32}$ eV, 
 consistent with the results observed with \xmm within errors. 

 {\bf \noindent Optical to X-ray SED.} 
Figure 3 shows the UV and optical SED of the transient for epochs where the quasi-simultaneous 
\swift UV and OGLE I-band observations are available, along with the best-fitting 
blackbody curves. 
The host UV flux is estimated from the host SED fitting 
{(see Methods: UV to optical SED analysis)} and subtracted from the observed emission. 
The blackbody fitting results indicate that the temperature appears roughly 
constant around $T\simeq16000$ K for the first $\sim40$ d after the optical discovery, 
and then rises to $T\simeq25000$ K over the next {$\sim$30 d whilst optical emission re-brightens}. 
Fitting to only the \swift UV data shows that the temperature does not to increase 
further for the rest of the epochs (Supplementary {Figure 4}). 
On the other hand, the radius remains at $\sim2\times10^{15}$ cm for the first $\sim40$ d, 
on the high end of the radius range observed for TDEs\cite{Hung2017}, after which it decreases by 
a factor of 3.  
Also shown in Figure 3 is the quasi-simultaneous X-ray spectrum observed with \xmm at $\sim$ 140 d 
and the best-fitting blackbody model. 
It can be clearly seen that while the X-ray temperature is an order of magnitude higher, 
it is not enough to explain the observed UV/optical emission, suggesting physically 
distinct emission components at the two bands arising probably from
different locations. 
{The blackbody radius inferred from the XMM observations 
 is $r_{\rm bb}\sim10^{12}$ cm, comparable to the Schwarzschild radius ($R_{\rm s}=2GM/c^2$) 
 for a black hole mass of $1.6\times10^{6}$\msun (\cite{Wyrzykowski2017}, and Supplementary Note 1). 
 This suggests that the soft {X-ray emission} originates from a compact accretion disk. 
 The origin for the UV/optical emission is not clear yet, and will be discussed in the next Section. 
}

The evolution of the X-ray luminosity {with respect to} UV/optical luminosity 
is displayed in Figure 4 (red symbols). In comparison with the X-ray brightening in other optical TDEs, 
{the evolution for \src presents a sharp increase at $\sim$150 days 
by about two orders of magnitude within only $\sim$2 weeks. 
The same trend can be seen from the X-ray luminosity evolution. 
Note that due to the lack of enough data points to verify the actual rise time, it can not be 
determined whether the rise to peak time of the X-ray emission for other optical TDEs 
is as dramatic as \srcs~(Supplementary Note 3 and Figure 5).  

}





\section{Discussion}
 \src is only the seventh optically discovered TDE with a detection of 
 bright X-ray emission ($L_{\rm X}/L_{\rm opt}\sim1$ close to the X-ray peak) by \swift and \xmm 
 within a few months since its discovery,  
 and the first source that has a resolved rise-to-peak in both X-ray and optical bands\cite{van2016, Gezari2017, Wevers2019, van2020}. 
 The X-ray emission exhibits a delayed brightening roughly $\sim$140 days  
 with respect to the optical emission which is also unique among optically discovered TDEs. 
 Many recent numerical studies have shown that the infalling stellar debris stream will 
 undergo self-intersections as a consequence of relativistic apsidal precession 
 \cite{Shiokawa2015, Bonnerot2017, Lu2020}, where 
 optical/UV emission could be produced because of shock heating. 
 Following the stream self-interactions, the debris spreads inward and gradually circularizes 
to form an accretion disk on the free-fall timescale.   
 {Within this picture, there will be a time delay between the debris self-crossing and onset of 
 disk formation, possibly explaining the observed delay of the X-ray emission in \srcs. However, 
 recent simulations of realistic disk formation\cite{Bonnerot2020} suggest that the shock heating rate of the initial 
 self-intersections near the apocenter radius might be much weaker than that required to power the 
 optical emission ($\sim10^{44}$ \erg), hence it appears not enough to account for observations.  
 Although the simulations show that self-intersections taking place close to pericenter can produce secondary 
 shocks with high enough heating rate, in this case the debris has reached a significant level of circularization, 
 leading to rapid formation of disk, which seems contrary with the early non-detection of X-ray emission. 
 Therefore, the scenario that the late time X-ray brightening in \src is due to delayed onset of disk formation seems disfavored.
 }

 {It has been proposed\cite{Metzger2016} that if the majority of 
      falling-back debris becomes unbound in a dense outflow, the X-ray radiation from the inner accretion 
      disk will be initially blocked, and may escape at later times as the density and opacity of the 
      expanding outflow decreases\cite{Wevers2019}. 
      In the model, efficient circularization of the returning debris is assumed, resulting in 
      rapid onset of disk accretion. 
      The time scale for the ionization break out of {X-ray radiation} is found about several months for a black hole 
      of $M_{\rm BH}\sim10^6$\msun\cite{Metzger2016}, in agreement with the observed time delay of 
      X-ray brightening for \srcs. 
      In this case, the reprocessing by irradiated ejecta can produce the bright optical emission, for which 
      the radiative efficiency is high enough to naturally explain the bolometric output of most TDEs. 
      The optical emission at early times for \src is likely due to the reprocessing of the {X-ray radiation}. 
      This is further supported by the ratio of X-ray to optical luminosities (Figure 4) that is very close to one at 
      the peak. 
      However, if the rapid X-ray brightening is due to the ionization break out of disk emission, the reprocessing 
      scenario alone cannot explain the late time X-ray evolution either, which is characterized by multiple flux dips and flares.  
    
     Alternatively, if the reprocessing layer is moderately patchy, the Keplerian motion 
    could be invoked to explain the unusual X-ray variability observed in \src due to 
    changes in absorption along the line of sight\cite{van2020}. We can estimate 
    the crossing time for an intervening gaseous material orbiting outside the X-ray source as \
    \begin{equation}
    t_{\rm cross}=0.7(\frac{r_{\rm orb}}{\rm light-day})^{3/2}M_{6}^{-1/2}{\rm arcsin}(\frac{r_{\rm abs}}{r_{\rm orb}})~{\rm yr},   
     \end{equation}
     where $r_{\rm orb}$ is the orbital radius of absorber, $M_6$ is the BH mass in units of $10^6$\msun, and $r_{\rm abs}$ 
     is the projected size of the moving absorption gas that is assumed comparable to the orbital radius\cite{LaMassa2015}. 
     Assuming that the distance of the intervening gas is the same as the radius of the photosphere for the UV/optical emission, 
     which is $\sim1-2\times10^{15}$ cm ($\sim0.4-0.8$ light day) from the blackbody fittings, and a black hole mass of 
     $10^{6.2}$\msun\cite{Wyrzykowski2017}, we obtained a crossing time of $\sim$50-140 days. 
     This is comparable to the duration of X-ray non-detections ($t\sim140$ days, the period between the optical peak and 
     first appearance of X-ray emission), as well as the time interval of following X-ray flares {(Figure 1b)}.
     However, since the rise to peak time is short ($\sim$10 days), such a scenario requires extreme condition
     such as sharp column density transitions 
     near the gap of the intervening gas and so partial covering of the X-ray source
     to accord with the almost no spectral changes between the low and 
     high X-ray flux states. In addition, the obscuring gas is required to be dense enough to 
     remain opaque for more than one year to block {X-ray source} where the reprocessing efficiency is expected high. 
     In contrast, the optical emission displays an extended plateau between 150 and 300 days, with a flux comparable 
     to the host emission observed in the pre-flare phase, suggesting that reprocessing may be less efficient.  
     The lack of continuous UV observations prevents further investigation on how the UV emission evolved at this epoch,  
     which is crucial to constrain the obscuration scenario. 
     Given the limited dataset, we cannot fully exclude the possibility of that the X-ray behavior is due to the 
     presence of the variable absorption, as proposed for the TDE AT2019ehz\cite{van2020}.  
      
  }

   Although it is rarely seen in standard TDEs, the strong flux dip superposed on the overall decline 
   {appears to be} consistent with the model prediction of tidal disruption by a SMBH binary (SMBHB) system, 
   where the presence of a secondary perturber can cause gaps in the light curve\cite{Liu2009, Ricarte2016, Coughlin2017}. 
   Such a characteristic flux interruption in the light curve has been observed in SDSS J120136.02+300305.5 (J1201+3003), 
   the first candidate TDE by a SMBHB in a quiescent galaxy \cite{Saxton2012, Liu2014}. 
   We test this possibility by using the same model {proposed for J1201+3003\cite{Liu2014} with} $M_{\rm BH}=10^6$\msun~
   and $\theta=0.3\pi$, where $\theta$ is the inclination angle between the orbital plane of SMBH binary and 
   disrupted star. We find that the overall X-ray light curve for \src can be reproduced by the model of tidal 
   disruption by a SMBH binary. The fitting results are summarized in Table 3 and shown in Figure 5a as grey 
   dot-dashed lines. Despite the high number of free parameters, in comparison with the fitting results for 
   J1201+3003 (Table 3), the best-fit of the SMBHB model for \src suggests  
   relatively large eccentricities ($e \sim 0.4 - 0.6$) and penetration factors ($\beta \sim 3 - 6$), while the orbital period of SMBHB is 
   similar with $T_{\rm orb} \sim 150 ~ {\rm days}$. The constraints on mass ratio are quite 
   uncertain. Both major and minor merger in the models are consistent with the observation ($q \sim 0.05 - 0.9$). 
   Since no clear tidal features are observed in the optical imaging\cite{Wyrzykowski2017}, the minor merger like 
   that inferred in J1201+3003 may be more favored. 
   
   Note that while the X-ray spectra for both objects 
   are extremely soft without emission at energies above 2 keV, a single thermal blackbody model is not sufficient 
   to describe the data, requiring an additional blackbody component (Supplementary Note 4 and Figure 6). 
   { 
   We argue that such an excess component is unlikely related to the disk emission of secondary BH 
   in the SMBHB scenario. Given the BH mass of $10^5-10^6$\msun~for the secondary (Table 3), 
   the best-fit blackbody temperature is much higher than that expected from 
   standard disk model\cite{Miller2015}, especially for J1201+3003. 
   In addition, as the separation of two SMBHs in our model configuration is relatively large compared with 
   the semi-major axis of the most bound debris, most of materials will be accreted by the primary BH 
   and the contribution from secondary to the X-ray emission is low. 
   In fact, such an additional X-ray emission appears to be ubiquitous in the X-ray spectra 
   of optical TDEs presented in Figure 4, and is found to vary in flux similar to the primary blackbody component\cite{Kara2018, Liu2019}. 
   It is likely that the extra component originates from a transient corona that is synchronized with the formation of 
   accretion disk (Supplementary Note 5 and Figure 7). Accurate modeling the evolution of the extra X-ray emission, 
   beyond the scope of this paper, is necessary to understand its origin.

   }

  
   If the interpretation of the SMBHB is correct, we note that the best-fit model predicts 
   an episodic accretion for a period of $\sim$90 days after the first interruption, producing an X-ray 
   luminosity of $\sim10^{43}-10^{44}$\erg, which is at odds with the \swift non-detections at that 
   period from either individual or combined observations. 
   {The observed properties for \src may be broadly consistent with the reprocessing model\cite{Metzger2016}. 
   The stream collisions lead to the rapid formation of the accretion disk, where the circularization process is 
   efficient likely due to the presence of a secondary BH\cite{Ricarte2016}.   
   The early X-ray non-detections could be attributed to the obscuration by a dense column of gas from unbound outflow, 
   where the radiation heating produces the UV/optical emission. 
   The ionization break out could allow the escape of {X-ray photons} at later times, yielding a delayed X-ray emission for which 
   the subsequent evolution is dominated by the discrete accretion in the SMBHB system. 
   }
   We note that the properties of ASASSN 14-li, the only TDE that is both X-ray and 
   optically-luminous since discovery\cite{van2016}, can be unified by the reprocessing scenario 
   if it is viewed along the direction with lower density of the ejecta 
   and so does the negligible time for ionization break out.   
    

  It is interesting to note that the optical I-band lightcurve of \src displays a second peak (rebrightening) 
   around 30 days after the initial peak. During the rebrightening phase, 
   the source also exhibits a possible variability in the UV bands. 
   While the nature of the rebrightening is under debate, 
   \cite{Wyrzykowski2017} has proposed the possibility of a binary BH on a tight orbit 
   to explain the optical variability in \srcs. 
   This is reminiscent of the TDE candidate ASASSN-15lh, for which a similar rebrightening in its 
   {optical/UV lightcurve has been observed and could be explained with a model of SMBHB with extreme mass 
   ratio ($q=0.005$)\cite{Coughlin2018}.   
   Strictly speaking, the SMBHB interpretation for ASASSN-15lh is qualitative, as the simulations show only the 
   evolution in the accretion rate for the secondary BH with a mass of $5\times10^5$\msun. In this case, the expected 
   luminosity from Eddington accretion is by an order of magnitude less that observed UV/optical luminosity ($\sim10^{45}$\erg), 
   making it unlikely that the UV/optical emission originates directly from the accretion disk.   
    The rebrightening can alternatively be explained as reprocessing of X-ray radiation into UV/optical emission 
    by delayed disk accretion\cite{Leloudas2016} onto a single SMBH, or ionization break out due to a sudden change in 
    the ejecta opacity\cite{Margutti2017}. 
    For the latter case, since the model is only sensitive to the UV emission, it seems difficult to explain the 
    optical plateau phase at longer wavelengths\cite{Leloudas2016}. 
    Hence the ionization break out is impossible to account for the optical rebrightening in \srcs, and the reprocessing of accretion 
    luminosity remains the most likely scenario. Since the rebrightening phase appears relatively short-lived, the 
    reprocessing scenario requires fluctuations in the mass accretion rate.  
    This is not inconsistent with the SMBHB model because of the presence of many accretion islands in the early times, 
    as shown in {Figure 5a}.
     Unfortunately, as we lack enough data points after the rebrightening phase in the I-band light-curve, 
    it cannot be determined whether the optical emission will display further variability or decay smoothly, 
    which is crucial to test the SMBHB scenario. 
    }




  {We conclude that the overall properties of \src could be accounted for by the stellar tidal 
   disruption by a candidate SMBHB.   
   The delayed brightening in the X-ray emission as well as the multiple flux dips during the decay phase 
   are in agreement with a SMBHB model with a mass of $10^6$\msun~for the primary BH, mass ratio of 0.25 and 
   orbital period of 150 days (Table 3). 
   In comparison with the prediction by the SMBHB model, the X-ray non-detections in the early phase 
   could be attributed to the obscuration by a dense column of gas, perhaps from unbound outflow. 
   This implies that the reprocessing may be a viable mechanism to explain the UV/optical emission 
   at the same epoch. 
   Ionization break-out allows for the escape of {X-ray photons}, resulting in the detectable X-ray emission at later times ($t\simgt140$ days). 
   A schematic illustration of this process is shown in {Figure 5b}. 
   }
   If our interpretation with the SMBHB model is correct, \src could be the second TDE candidate 
   with a SMBHB at its core revealed in the X-rays. 
   Upon final coalescence, SMBHB systems like the one in \src (and SDSS J1201+3003) are prime sources for future space-based 
   {GW missions like {Laser Interferometer Space Antenna}\cite{McGee20}.} 
   {Note that given the estimated GW inspiral time of $t_{\rm gw}\sim1.9\times10^7$ years\cite{Peters1963}, 
   it would be challenging to 
   detect the GWs from such SMBHB system in its current state of evolution. }
   {In synergy with {Large Synoptic Survey Telescope}\cite{LSST2009}, the future X-ray sky surveys such as 
   {extended Roentgen Survey with an Imaging Telescope Array}\cite{Predehl2020} and {Einstein Probe}\cite{Yuan2018} are 
   expected to detect similar TDEs more than one hundred\cite{Thorp2019},  }
   providing a powerful tool for studying the physics on how the 
   stellar debris evolves after disruption, and searching for {promising} candidates of milliparsec SMBHBs 
   that are still poorly explored. 
   
    \clearpage
    \newpage

    \section{Methods}

    \label{sec:data}

    {\bf \noindent X-ray data reduction.} 
    \src has two \xmm observations, which were performed on Jun 2016 and Nov 2016 
    ({XMM1} and {XMM2}), with an exposure of 15ks and 36 ks, respectively. 
    The \xmm data were reprocessed with the
    Science Analysis Software version 16.0.0, using the calibration
    files that are available up to 2018 December. 
    The time intervals of high background events were excluded by 
    inspecting the light curves in the energy band above 12 keV where the count rates for source are low. 
    Detailed spectral analysis was performed only on the data taken with PN\cite{Struder2001}, as it has a 
    much higher sensitivity, while the MOS\cite{Turner2001} data have been used to check for consistency when necessary. 
    The source spectra were extracted within a circular region centered at the source optical coordinate, with a radius of {35}\arcsec. 
    Background spectra were extracted from clean regions on the
    same chip using four identical circular regions to that of source. 
    We grouped the spectra to have at least 5 counts in each energy bin so as
    to adopt the $\chi^2$ statistic during the process of spectral fitting.  
    Since no hard X-rays are detected at energies above 2 keV, we performed spectral fittings in the 0.3-2 keV range 
    using XSPEC (version 12). All statistical errors given correspond to the 90\% confidence intervals for one interesting 
    parameter ($\Delta\chi^2=2.076$), unless stated otherwise.

    All \swift observational data were retrieved from the HEASARC data archive. 
    Details of the \swift observations can be found in Supplementary Table 1. 
    The calibration files are taken from that released on 2017 November 13. The XRT was operated in Photon Counting
    mode. We reprocessed the XRT event files with the task
    {xrtpipeline} (version 0.13.2). 
    For each \swift XRT observation, we used XSELECT which is part of HEASoft (FTOOLS 6.19) 
    to extract the source spectrum with a circular region of {40}\arcsec~radius. 
    Background spectrum was extracted from an annulus region centered on the source position,  
    with an inner radius of 60\arcsec~and outer radius of {110}\arcsec, respectively. 
    For the data taken in first 16 epochs (from 2016 Jan 19 to June 8), no X-ray signal was observed at the 
    location of \src in either individual or stacked images. The corresponding $3\sigma$ upper limit was estimated 
    from the background region with the spectral extraction task in the HEASoft package. 
    The source is not detected either on 2016 Nov 17 and the epoch between 2017 Feb 19 and Feb 23. 
    The X-ray upper limits as well as the detections from \swift observations are listed in Table 1. 
    The count rates were converted into flux using the WebPIMMS tool, assuming a blackbody model with 
    temperature of {$kT_{\rm BB}=60$ eV} modified by a Galactic HI column density of $N_{\rm H}=2.71\times10^{20}$ cm$^{-2}$. 

    {\bf \noindent Ultraviolet and optical data reduction.} UV imaging data of \src were obtained with the 
    \swift UVOT instrument in three filters: UVW1 (2600\AA), UVM2 (2246\AA) and UVW2 (1928\AA). 
    For the most recent observations performed on 2020 Jan 12, imaging data in three optical filters, 
    U (3465\AA), B (4392\AA) and V (5468\AA), are also available, which allow for better determining the 
    host emission. 
    We used the UVOT software task {\tt UVOTSOURCE} to extract the source counts from a region with 
    radius of 5\arcsec. The background is chosen from a source-free sky region with a radius of 
    40\arcsec. The UVOT count rates were then converted into AB magnitudes which are presented in 
    Supplementary Table 1. The OGLE-IV I-band photometric data are publicly available and can be 
    found in http://ogle.astrouw.edu.pl/ogle4/transients/. The host contribution to the I-band flux 
    was subtracted using the pre-discovery images as templates. 
    For the UVOT data, since no good pre-discovery reference images are available and 
    the photometric errors from the most recent observations are large, we used the model galaxy 
    spectrum from the SED fit to generate synthetic magnitudes at these wavelengths.
    We obtained the pre-flare measurements of \src from the NED database (https://ned.ipac.caltech.edu), 
    including NUV and FUV photometry from GALEX, optical b\_J photometry from the APM survey, 
    near-IR JHK$_{\rm s}$ photometry from the 2MASS, and mid-IR photometry at 3.6\um and 4.5\um from WISE.  
    In addition, we also used the UVOT photometry from the most recent observations taken nearly 1480 days 
    since optical outburst, for which the emission is likely dominated by the host component. 
    All the host photometric data are presented in Supplementary Table 2.   
    We fitted the above photometry of the host using the code {\tt FAST} and the 
    best-fitting model is shown in {Supplementary Figure 1 (and Note 1)}. 
    In the AB system, we obtained the host magnitudes of {UVW1 = 20.52$\pm$0.18 mag, UVM2 = 20.75$\pm$0.21 
    mag and UVW2 = 21.03$\pm$0.18 mag},  
    which were then subtracted from the UVOT measurements 
    to obtain the transient photometry {(and the errors on host flux were propagated)}. 
    In addition, all flux densities have been corrected for the Galactic extinction of $E$(B-V) = 0.018 mag. 
    Note that we did not corrected for an internal extinction by host, as the uncertainties on the best-fit extinction 
    given by {\tt FAST} are large ($A_{\rm V}=1.0^{+1.65}_{-0.3}$, 68\% confidence intervals).   

    {\bf \noindent Ultraviolet to optical SED analysis.} 
    We fitted a blackbody model ($B_{\nu}=\frac{2h\nu^3}{c^2}\frac{1}{e^{h\nu/kT}-1}$) to the host subtracted, extinction-corrected photometric data 
    from the \swift UVOT observations, to put constrains on the luminosity, temperature 
    and radius evolution of UV and optical emission. Uncertainties on the above 
    parameters were derived using Monte Carlo simulations, in which the observed 
    fluxes were randomly perturbed with amplitude by assuming Gaussian noise according to 
    the photometric errors.  This procedure was repeated 1000 times.  
    The error bars on each parameter were then derived from the 16th 
    and 84th percentiles of the distribution of the corresponding values 
    obtained in the simulations. Using the best-fit temperature and rest-frame monochromatic 
    UV luminosity at each epoch, we estimated the blackbody radius from 
    $L_\nu=\pi B_\nu\times4\pi R_{\rm BB}^2$ and took blackbody luminosity 
    from $L_{\rm BB}=\sigma T^4 \times4\pi R_{\rm BB}^2$ as the integrated 
    luminosity of UV and optical emission.  The evolution of blackbody luminosity, 
    temperature and radius are presented in the Supplementary {Figure 4}.

   \begin{addendum}
   \item [Data availability]  
	
	   {Source data for the observations taken with \xmm and \swift are available through 
	the HEASARC online archive services (https://heasarc.gsfc.nasa.gov/docs/archive.html). 
         Optical imaging data at I-band are publicly available through the website of 
	 the OGLE-IV Transient Detection System (http://ogle.astrouw.edu.pl/ogle4/transients/2017a/transients.html). 
	 The authors can provide other data that support the findings of this study upon request.  
	   }
   \end{addendum}

  \begin{addendum}
	     \item [Code availability]
		     {{The Science Analysis Software used to reduce the \xmm data is publicly available at  
		          https://www.cosmos.esa.int/web/xmm-newton/download-an. 
               The X-ray analysis softwares, XSPEC and XSELECT, are part of HEASoft, which can be found 
	       at 
	       https://heasarc.gsfc.nasa.gov/
	       docs/software/heasoft.
	       The \swift data analysis software (xrtpipeline and UVOTSOURCE)  
	       can be found at https://swift.gsfc.nasa.gov/analysis/. 
	       WebPIMMS tool is available at 
	       https://heasarc.gsfc.nasa
	       .gov/cgi-bin/Tools/w3pimms/w3pimms.pl. 
	       The code used to model the UV-to-MIR SED is accessible through github (https://github.com/jamesaird/FAST). 
             The other codes that support the plots within this article are available from the 
     authors upon reasonable request.}} 
  \end{addendum}



\renewcommand{\refname}{References}

\clearpage

\begin{addendum}
\item [Acknowledgements]
This research made use of the HEASARC online data archive services, supported by
NASA/GSFC. We thank the \swift and \xmm observatories, and the OGLE-IV Transient Detection System 
(OTDS) for making the data available. {X.S. thanks Zuozifei Song for producing the 
	schematic diagram in Figure 5.  }  
This work is supported by Chinese NSF through grant Nos. 11822301, 11833007, {U1731104 and 11421303. 
S.L. is grateful to the Key International Partnership Program of the Chinese Academy of Sciences (CAS) 
(No. 114A11KYSB20170015), and the Strategic Priority Research Program (Pilot B) `Multi-wavelength gravitational wave universe' 
of CAS (No. XDB23040100). 
} 

\item[Author Contributions] 
X.S. performed the data analysis and prepared the paper. 
W.Z., L.D. reduced the \xmm and \swift spectral and photometry data. 
S.L. performed the X-ray light curve fitting with the SMBHB TDE model. 
N.J. and L.S. commented the optical and X-ray data analysis, and 
contributed to the UV-optical spectral energy distribution fitting.   
T.W., F.L., Z.Y., F.X. and R.S. contributed to the overall interpretation of the results. 
All the authors joined the discussion at all stages. 

\item[Competing Interests]
{The authors declare no competing interests. }

\end{addendum}

\renewcommand\figurename{\footnotesize \textbf{Figure}}
\renewcommand{\thefigure}{\textbf{\footnotesize \arabic{figure}}}

\newpage
\begin{figure*}[t]
\centering
 \includegraphics[scale=1.0]{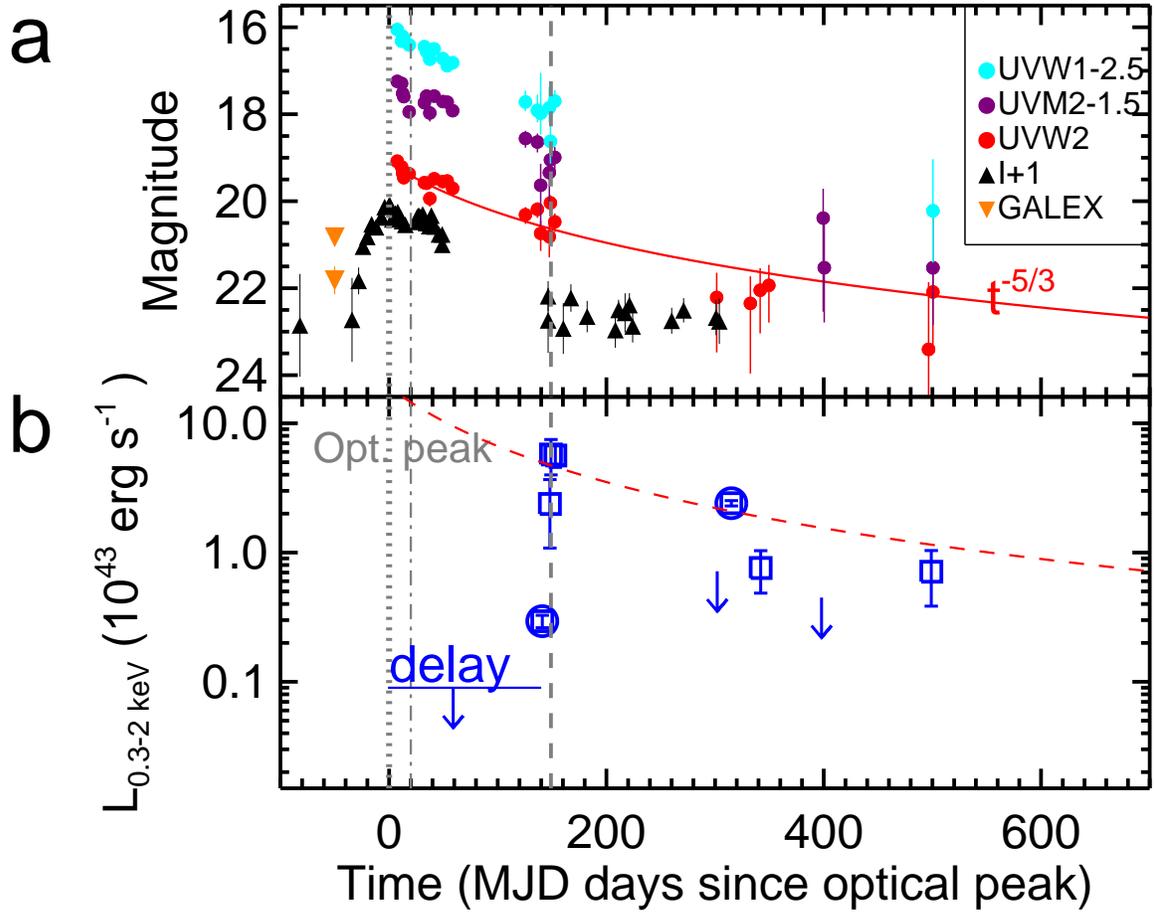}
 \caption{
 {UV-optical and X-ray light curve of \srcs.}
{\bf a} The host-subtracted, Galactic dust extinction-corrected UV and optical light curve. 
The \swift UVW1, UVM2 and UVW2 photometric data are 
shown in cyan, dark purple and red, respectively. We also plot the pre-flare GALEX UV data for comparison 
(orange). The optical I-band data are shown in black triangles. The light curves are shifted 
by constants for clarity as noted in the legend. All magnitudes are in the AB system. 
Error bars represent 1$\sigma$ uncertainties due to photometric errors. 
The red line represents the fit to the UVW2 data with a 
canonical $n=-5/3$ power-law decay, {assuming the same peak time as optical one\cite{Wyrzykowski2017} that 
is marked with gray dotted line. 
The dot-dashed line represents the approximate time of optical re-brightening, $t\approx20$ days with respect to the initial peak. 
The dashed line marks the time for the X-ray peak. 
Note that the true peak time might be slightly later, as no further observations are performed between the current peak and 
the non-detection at $t=302$ days. }
{\bf b} The follow-up X-ray observations from \swift (square) and \xmm (circle). 
{The 3$\sigma$ upper limits on the flux for non-detections (see Table 1 and text) are shown with downward pointing arrows. }
The X-ray emission shows a clear time delay with respect to optical by $\sim$140 days, and brightened to 
peak within only $\sim$2 weeks. All the X-ray luminosities are corrected for the absorption. 
{Error bars represent 1$\sigma$ uncertainties calculated using Monte Carlo simulations provided in XSPEC. }
{The red dashed line corresponds to the best-fit to UVW2 data in the upper panel, but scaled to the UV/optical blackbody 
luminosity at $t=153$ days (Supplementary Figure 4). 
}}
\label{fig:youngsed}
\end{figure*}

\newpage
\begin{figure}[tpb]
\centering
\includegraphics[scale=0.39,angle=0]{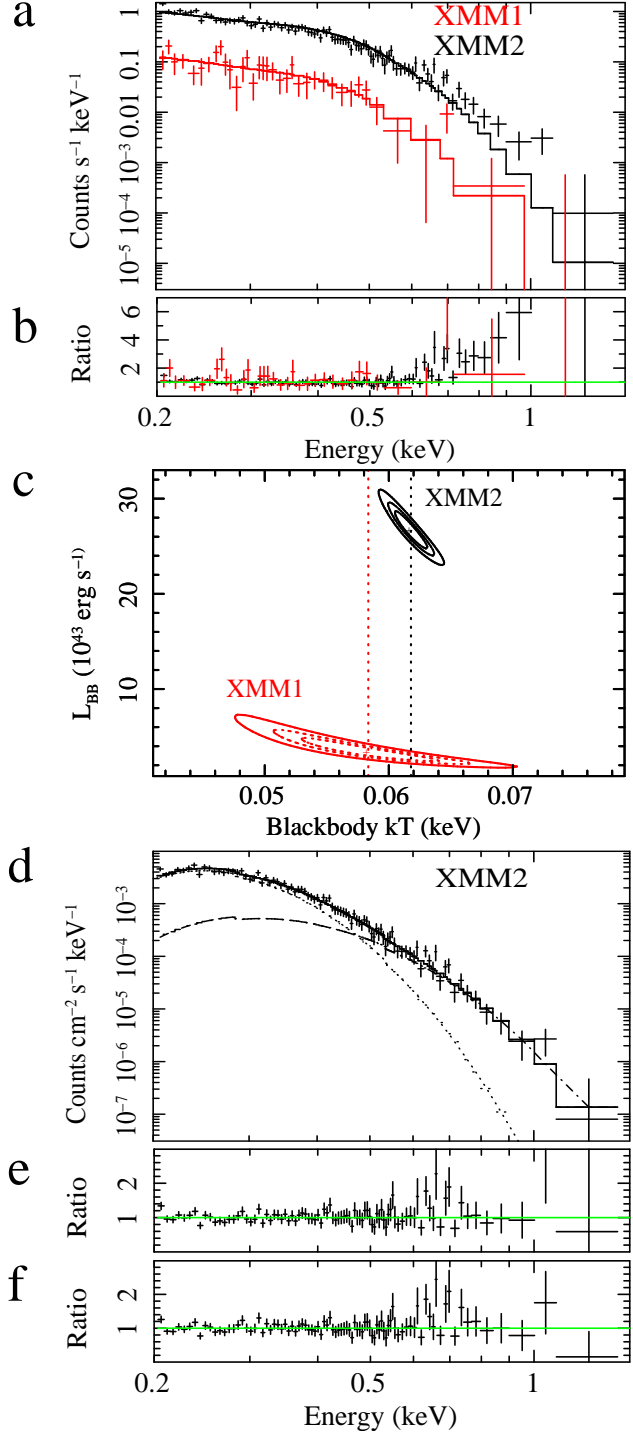}
 \caption
 {
 {Fittings to the X-ray spectra of \src from \xmm observations at two epochs.}
 {\bf a} \xmm PN spectra for \src along with the best-fitted single blackbody model. 
  {Error bars represent 1$\sigma$ uncertainties calculated using Poisson statistics.} 
 The data observed on June 9 2016 (XMM1) are shown in red, while that obtained on Nov 30 2016 
 (XMM2) shown in black. The corresponding data to model radios are shown in {\bf b}. 
 {\bf c} {shows the joint 68\%, 90\% and 99\% confidence contours of the 
 	 blackbody temperature versus luminosity for the two observations. 
         The vertical dotted lines represent the best-fitted temperatures. }
	 {{\bf d} shows the best-fitting result by including an additional blackbody component 
	 to account for the excess emission at above $\sim$0.7 keV for the XMM2 data which is shown in dashed line, 
	 and {\bf e} is the corresponding data/model ratio. 
	 {\bf f} the same as {\bf e}, but for the data/model ratio from spectral fitting by using 
         an additional powerlaw component to describe the excess emission. }
}
\end{figure}

\newpage

\begin{figure}[tp]
\centering
\includegraphics[scale=0.93,angle=0]{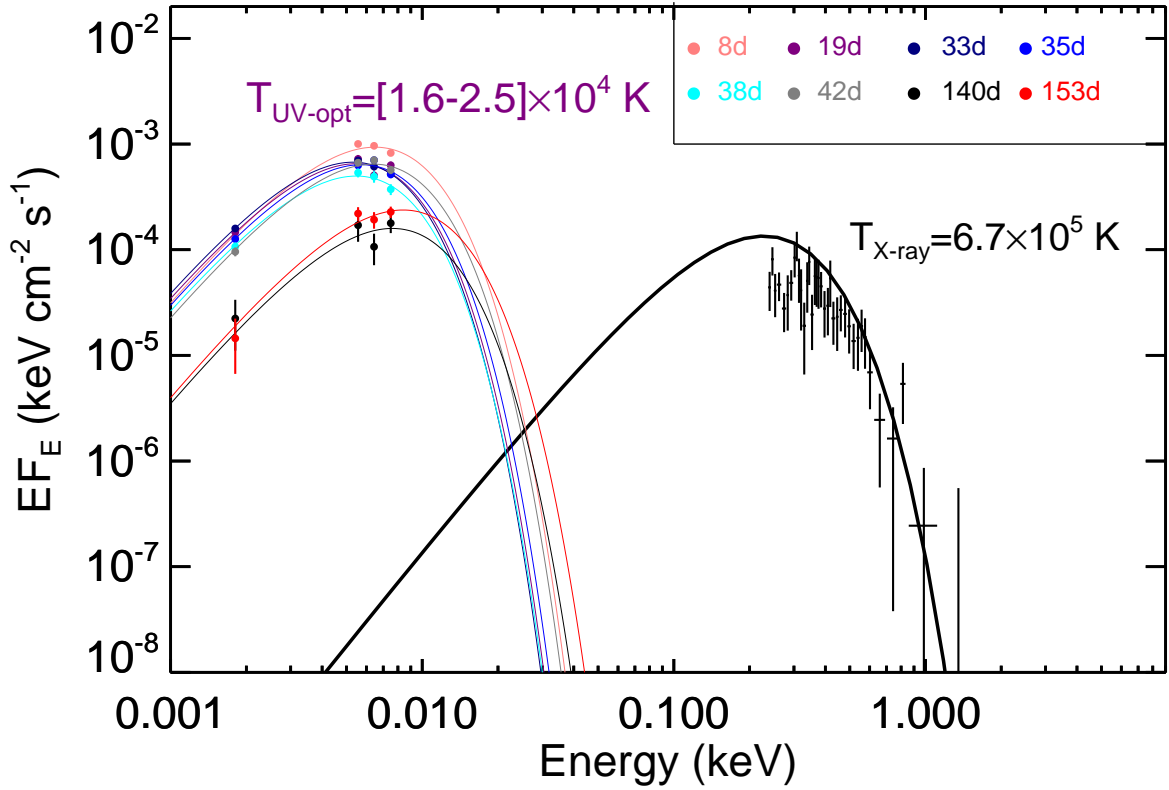}
 \caption
 {
 {Evolution of the UV to optical SED of \src at different epochs}. 
 The corresponding best-fitting blackbody models are shown in different colors {(noted in the legend), 
  with temperatures in the range $[1.6-2.5]\times10^4$ K. 
 The epochs refer to the time relative to the initial optical peak, in units of days (d).
 The quasi-simultaneous X-ray spectrum observed with \xmm (XMM1) about 140 days after the initial optical peak is shown in black, }
 which can be fitted with a {single blackbody model of $kT_{\rm BB}=58$ eV ($T_{\rm X-ray}=6.7\times10^5$ K)} modified by Galactic absorption (see text). 
 {Error bars represent 1$\sigma$ uncertainties that are same as that in Figure 1 and 2.}
 The unabsorbed blackbody model corrected for Galactic absorption is shown with black curve.  
 Although with a much higher temperature, the best-fit blackbody to the X-ray data severely underpredicts 
 the UV and optical flux, suggestive of different physical origins.  
 }
\end{figure}

\newpage

\begin{figure}[]
\centering
\includegraphics[scale=0.93,angle=0]{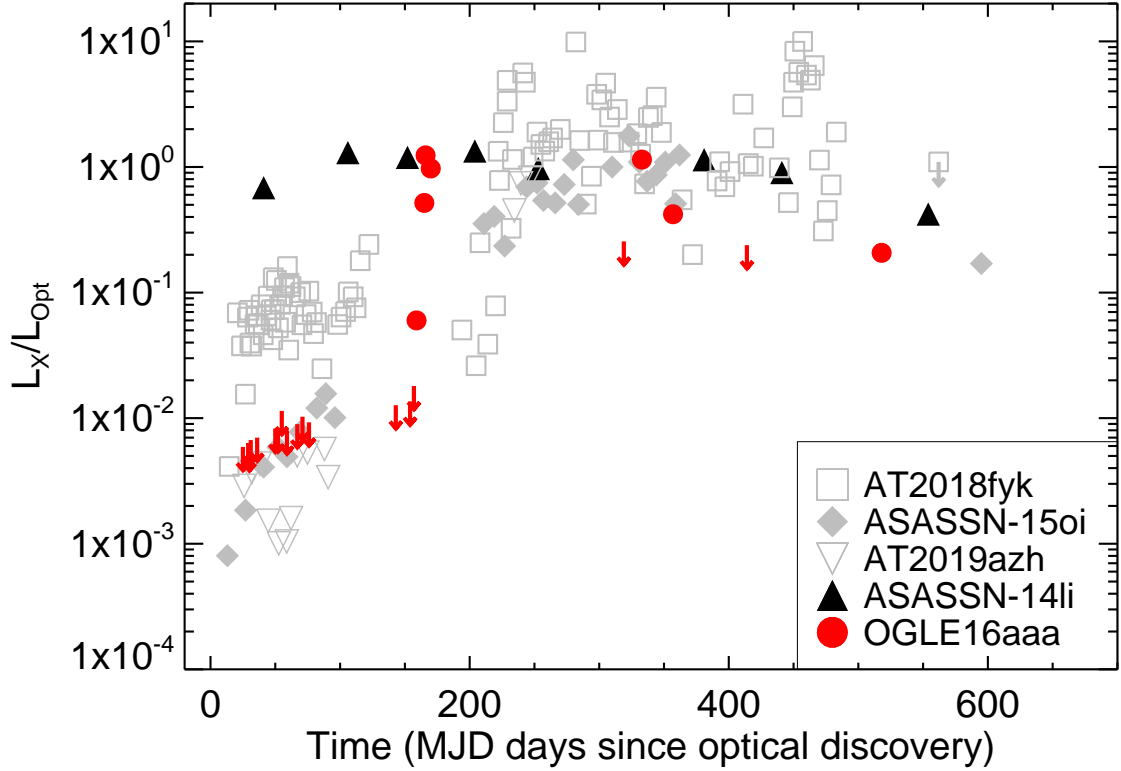}
 \caption{
 {Comparison of \src with other TDEs in the evolution of optical to X-ray luminosity ratio.} 
 The optical luminosity refers to the integrated blackbody luminosity from 
 the SED fittings to the \swift UVOT photometry. 
 {For epochs where only UV observation at one band is available ($t=[300,490]$ days), we extrapolated 
 the UV luminosity to the blackbody luminosity by assuming a constant temperature evolution since $t=150$ days.}
 For the X-ray non-detections, 
 the corresponding 3$\sigma$ upper limits are given. 
 {Red symbols represent the luminosity ratios of \src at different epochs. 
 For comparison, we also plotted the ratios for other X-ray bright TDEs from \cite{Gezari2017, 
 Wevers2019, Liu2019}, as noted in the legend.} ASASSN-14li\cite{van2016} is the only TDE that shows 
 nearly constant optical to X-ray luminosity ratios since discovery. While \src appears 
 to follow the evolution trend as other TDEs, the time of rise to peak is distinct. 
 }
 \end{figure}
\clearpage

\begin{figure}[]
\centering
\includegraphics[scale=0.67,angle=0]{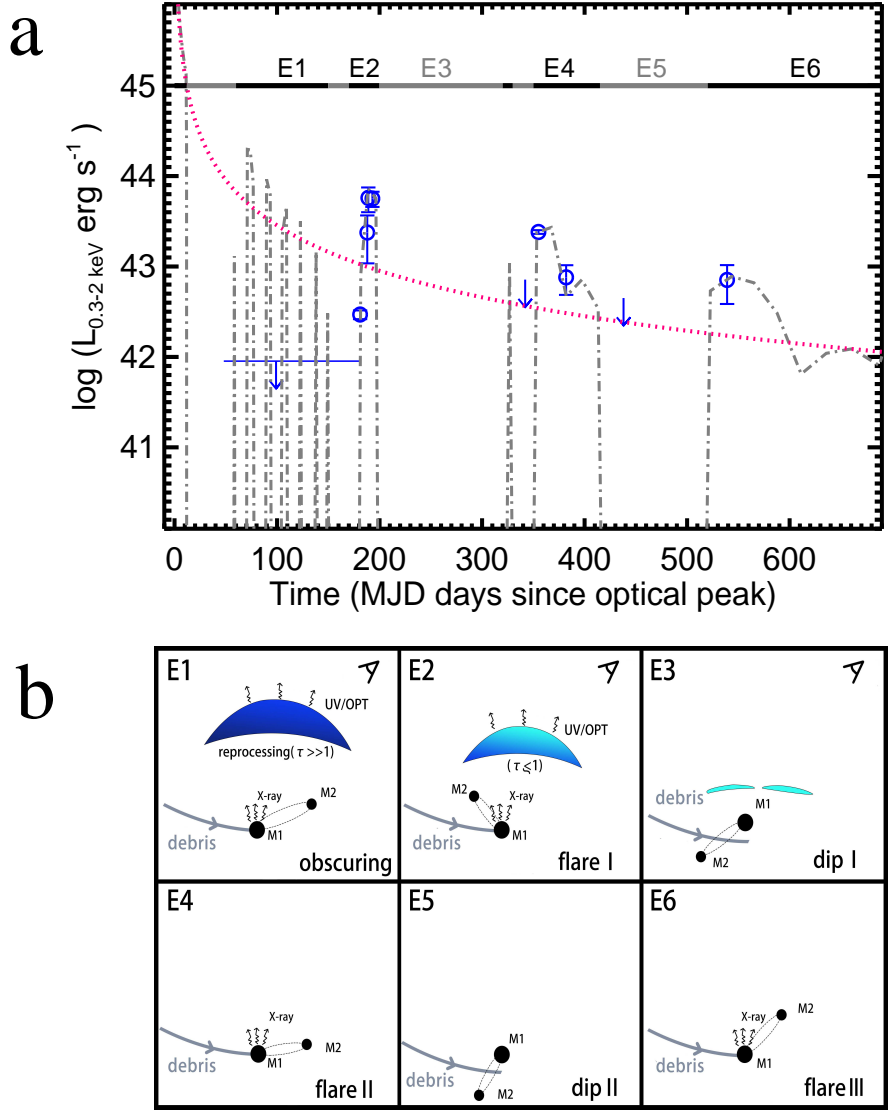}
 \caption{
 {X-ray luminosity evolution of \src modeled with the tidal disruption by binary black hole.}
 {\bf a} The observed 0.3--2 keV luminosity of \src at different epochs, shown with blue symbols.  
 {Error bars correspond to 1$\sigma$ uncertainties, as that shown in Figure 1.
  The 3$\sigma$ upper limits are adopted for X-ray non-detections and shown with downward pointing arrows.
 Simulated light curve is shown in grey dot-dashed line, for the tidal disruption by binary black hole in the observer frame for \src 
 (Table 3). }
 Since there is a time difference between the fallback particles and radiation 
 in simulations, which is 40 days from the best-fit, the observed light curve 
 is shifted rightward to match the simulated one. 
 The magenta-red dotted line represents the canonical $t^{-5/3}$ decay law. Observational data are from 
 Table 1. 
 {The horizontal thick lines represent different stages of X-ray evolution,} as illustrated in {\bf b}. 
 {\bf b} A schematic illustration of the evolution of X-ray emission. 
E1: The X-ray emission from the inner accretion disk of primary BH is initially obscured 
by a thick reprocessing layer, by which the X-ray radiation is absorbed and re-emitted 
at UV/optical wavelengths; E2: Ionization radiation results in a change in the opacity 
at later times, and direct escape of X-ray radiation (flare I). The reprocessing layer recedes 
inwards as $\tau\simlt1$; E3: Due to the perturbation of 
secondary BH, the infalling stream debris misses the accretion radius, causing interruptions 
in the light curve (dip I); E4: The accretion continues and X-ray flare occurs (flare II); 
E5: The infalling debris begins to miss again (dip II); E6: Beginning of another stage with accretion (flare III).
 }
 \end{figure}
\clearpage

 \renewcommand{\tablename}{Table}
 \setcounter{table}{0}
\begin{table*}
\footnotesize
\title{ Table 1. X-ray Observations of \src}\\
\begin{center}
\begin{tabular}{cllcccc} 
	\hline 
	\hline
Telescope & obsID & obs. date & days  &  exposure  & counts rate & flux	 \\
          &       &           &       &   sec &  $10^{-2}$ cts/s &  $10^{-13}$ \ergs \\
\hline
Swift-XRT & 00034281001--- & 2016 Jan 19 --- & 0-141 & 30160 &
$\textless0.023$ & $\textless0.12$ \\
                &   16                   &     2016 June 8                &       &
                     &                      & \\
XMM-PN & 0790181801 & 2016 June 9 & 141 & 10550 & $1.01\pm0.11$ & $0.40\pm0.05$ \\
Swift-XRT & 00034281018 & 2016 June 16 & 148 & 619 & $0.60\pm0.32$ & $3.15\pm1.72$ \\
Swift-XRT & 00034281019 & 2016 June 17 & 149 & 797 & $1.44\pm0.44$ & $7.59\pm2.31$ \\
Swift-XRT & 00034281020 & 2016 June 21 & 153 & 1974 &$1.42\pm0.27$ & $7.46\pm1.42$ \\
Swift-XRT & 00034281021 & 2016 Nov 17 & 302 & 1684 & $\textless0.18$ & $\textless0.94$ \\
XMM-PN & 0793183201 & 2016 Nov 30 & 316 & 21830 & $9.37\pm0.21$ & $3.18\pm0.15$ \\
Swift-XRT & 00034281024---& 2016 Dec 18--- & 333-350 & 4842
& $0.19\pm0.07$ & $1.01\pm0.36$ \\
              &       26               &     2017 Jan 4                &       &
                     &                      & \\
Swift-XRT & 00034281027--- & 2017 Feb 19--- & 397-401 & 2665
& $\textless0.11$ & $\textless0.59$ \\
           &         29             &     2017 Feb 23                &       &
                     &                      & \\
Swift-XRT & 00034281030---& 2017 May 31--- & 497-501 & 3014
& $0.18\pm0.08$ & $0.94\pm0.42$ \\
           &        31              &     2017 June 4                &       &
                     &                      & \\
Swift-XRT & 00034281032 & 2020 Feb 9 & 1481 & 1903 & $\textless0.16$ & $\textless0.84$ \\		     
\hline
\end{tabular}
\end{center}
{\bf Note--}
For non-detections in either individual epochs or combined data, the corresponding 3$\sigma$ upper limits on 
the counts rate and flux are given. 
The X-ray counts rate and flux is in the 0.3--2 keV, respectively. The X-ray flux  
is corrected for the internal and Galactic gas absorption. 
\end{table*}

\begin{table*} 
\footnotesize
\title{}Table 2. Spectral fitting results for the X-ray data\\
\begin{center}
\begin{tabular}{llllccc}
\hline \hline
 Model component & Parameter  &  XMM1 & XMM2 \\
            & &  2016 June & 2016 Nov\\ 
 \hline
Model 1&   \\
 Blackbody & $kT_{\rm BB}$ (eV)  &  58$\pm$5   &   60$^{+9}_{-11}$              \\
Power-law & $\Gamma$      &  $\dots$ & 6.4$\pm$0.5  \\
 Neutral absorber  & $N_\mathrm{H}$~($\times10^{22}$\nh)& $<$0.06 & $<$0.23  \\
\hline
Model 2& \\
Blackbody1 & $kT_{\rm BB1}$ (eV)  &    $\dots$ &   51$^{+5}_{-8}$              \\
Blackbody2 & $kT_{\rm BB2}$ (eV)  &    $\dots$ &   90$^{+20}_{-13}$          \\
Neutral absorber  & $N_\mathrm{H}$~($\times10^{22}$\nh)& $\dots$ & $<$0.03  \\
\hline
Statistics  & $\chi^2/$dof &   25.6/35   & 80.6/86 (82.2/87)  \\
 \hline
\end{tabular}
\end{center}
{\bf Note--}
The $\chi^2$ statistics in the parentheses is for Model 2. 
\end{table*}

\begin{table*}
\footnotesize
\title{} Table 3: SMBHB model parameters for \src {and J1201+3003}.\\
\begin{center}

\begin{tabular}{lllccccccc}
\hline\hline 
Parameter & \src & J1201+3003 \\
\hline
BH mass (\msun) & $10^6$ & $10^7$ \\
Eccentricity $e$ & 0.4 [0.4, 0.6] & 0.3 [0.1, 0.5] \\
Penetration factor $\beta$ & 4.5 [3.0, 6.0] & 1.3 [1.3, 1.6] \\
Mass ratio $q$ & 0.25 [0.05, 0.9] & 0.08 [0.04, 0.09] \\
Orbital period $T_{\rm orb}$ (days)& 150 [140, 160]  & 150 [140, 160] \\
Initial phase $\phi$ & 1.7$\pi$ & 1.5$\pi$ \\
\hline
\end{tabular}
\end{center}
\end{table*}

\end{document}